\documentclass[11pt,reqno,a4,onecolumn,twoside,final,spanish]{article}

\usepackage{amsthm,amsmath,amssymb,latexsym,amsfonts,amscd} 
\usepackage{latexsym}
\usepackage{url}
\usepackage[T1]{fontenc}
\usepackage{ae}
\usepackage{aecompl}
\usepackage[small,sc,hang,bf]{caption} 
\usepackage{graphicx,color,ae,fancyvrb}
\usepackage{pst-all}
\usepackage{subfig}
\usepackage{float} 
\usepackage{verbatim}
\usepackage{enumerate}
\usepackage{array}
\usepackage{longtable}
\usepackage{multicol}
\usepackage{cancel}
\newcommand*\chancery{\fontfamily{pzc}\selectfont}

\newlength\mytemplen
\newsavebox\mytempbox

\makeatletter
\newcommand\mybluebox{%
    \@ifnextchar[
       {\@mybluebox}%
       {\@mybluebox[0pt]}}

\def\@mybluebox[#1]{%
    \@ifnextchar[
       {\@@mybluebox[#1]}%
       {\@@mybluebox[#1][0pt]}}

\def\@@mybluebox[#1][#2]#3{
    \sbox\mytempbox{#3}%
    \mytemplen\ht\mytempbox
    \advance\mytemplen #1\relax
    \ht\mytempbox\mytemplen
    \mytemplen\dp\mytempbox
    \advance\mytemplen #2\relax
    \dp\mytempbox\mytemplen
    \colorbox{myblue}{\hspace{1em}\usebox{\mytempbox}\hspace{1em}}}

\makeatother

\usepackage{natbib}

\usepackage{hyperref}

\usepackage{geometry}
\usepackage{calc}
	
\makeatletter
\renewcommand{\section}{\@startsection{section}{1}{0pt}{-3ex plus -1ex minus 0ex}{2ex plus 0ex}{\bf}}
\makeatother

\makeatletter
\renewcommand{\subsection}{\@startsection{subsection}{1}{0pt}{-2ex plus -1ex minus 0ex}{2ex plus 0ex}{\bf}}
\makeatother

\theoremstyle{definition}

\theoremstyle{remark}

\begin{document}
\renewcommand{\tablename}{Tabla}
\renewcommand{\figurename}{Figura}
\noindent

\begin{flushleft}
\textsl {\chancery  Memorias de la Primera Escuela de Astroestad\'istica: M\'etodos Bayesianos en Cosmolog\'ia}\\
\vspace{-0.1cm}{\chancery  9 al 13 Junio de 2014.  Bogot\'a D.C., Colombia }\\
\textsl {\scriptsize Editor: H\'ector J. Hort\'ua}\\
\href{https://www.dropbox.com/sh/nh0nbydi0lp81ha/AACJNr09cXSEFGPeFK4M3v9Pa}{\tiny {\blue Material suplementario}}
\end{flushleft}



\thispagestyle{plain}\def\@roman#1{\romannumeral #1}



\begin{center}\Large\bfseries Estudio de geod\'esicas nulas en un espacio axialmente sim\'etrico. \end{center}
\begin{center}\normalsize\bfseries Study of null geodesic in an axially symmetric space \end{center}

\begin{center}
\small
\textsc{Jonathan Pineda\footnotemark[1]}
\textsc{Leonardo Casta\~neda\footnotemark[2]}
\textsc{Juan Manuel Tejeiro\footnotemark[3]}
\footnotetext[1]{Universidad Nacional de Colombia. E-mail: \url{japinedag@unal.edu.co}}
\footnotetext[2]{Universidad Nacional de Colombia. E-mail: \url{lcastanedac@unal.edu.co}}
\footnotetext[3]{Universidad Nacional de Colombia. E-mail: \url{jmtejeiros@unal.edu.co}}

\end{center}

\noindent\\[1mm]
{\small
\centerline{\bfseries Resumen}\\
Desde el surgimiento de la teor\'ia de la Relatividad, se han encontrado una gran varie\-dad de soluciones en torno a las 
ecuaciones de campo de Einstein, dependiendo del tipo de espacio-tiempo. Entre ellas, se encuentra el espacio axialmente 
sim\'etrico que describe los agujeros negros rotantes. Para entender m\'as acerca de estos objetos astrof\'isicos se realiza 
un trazado de rayos, que es principalmente  una simulaci\'on de la trayector\'ia que siguen los fotones desde el agujero negro
hasta el observador y debido a que  son millones de fotones que cada segundo recibe el observador, se necesita una herramienta
computacional que permita resolver este problema para la formaci\'on de las  im\'agenes alusivas a un agujero negro
y obtenidas a trav\'es del uso de c\'odigos num\'ericos.  Estos \'ultimos,  permiten una perspectiva gr\'afica de cada p\'ixel generado por las
trayector\'ias seguidas por el fot\'on en el espacio-tiempo,  con lo que se puede analizar variables y caracter\'isticas del agujero negro. 
Con la ayuda del programa \textit{YNOGK-GEOKERR},  se pueden generar las im\'agenes del agujero negro y 
tambi\'en las soluciones de las  ecuaciones de movimiento de este sistema astrof\'isico. Con este
art\'iculo se prentende realizar un estudio sobre este tipo de espacio-tiempo y adem\'as se muestran  los resultados
obtenidos por la simulaci\'on de un agujero negro con esta rutina.\\

{\footnotesize
\textbf{Palabras clave:}
Trazado de Rayos, Separabilidad de Hamilton Jacobi, Coordenadas de Boyer-Lindsquist.
\\
\noindent\\[1mm]
{\small
\centerline{\bfseries Abstract}\\
Since the development of Relativity theory, a variety of solutions have been found around Einstein field equations, depending  on  the time-space.
For instance, we can find the axially symmetric space  which describes a rotanting black hole. To understand this astrophysics objects, we must do a ray
tracing,  which is a  simulation of photon's trajectory from the black hole to observer. Since are  millions of photons recievied by the observer, 
we need a computational tool to resolve this problem  for generating  the image of  black hole. Thanks  to the  
\textit{YNOGK- GEOKERR} code, we can visualise the black holes characteristics and  the solutions of the equations of motion using separability and interior point methods.
With this article we want to study  this type of space and also to show the results for black hole simulations with this routine.   
\\

{\footnotesize
\textbf{Keywords:}
Ray tracing, Hamilton Jacobbi separability, Boyer-Linquist coordinates.\\
}

\section{Introducci\'on}
En este art\'iculo estudiamos la soluci\'on de las ecuaciones de campo de Einstein y la obtenci\'on de las ecuaciones de movimiento
para un espacio axialmente sim\'etrico (espacio de Kerr) de un agujero negro rotante. Su inter\'es esta en hacer 
cambios de coordenadas para solucionar dichas ecuaciones. El objetivo principal de este trabajo es el uso
de herramientas computaciones que permiten solucionar este tipo de problemas  de manera m\'as eficiente y  haciendo uso de simulaciones que generan
una perspectiva distinta sobre los distintos par\'ametros del problema.\\
A partir de la relatividad general, donde se deduce la ecuaci\'on de la geod\'esica, que describe la trayector\'ia 
del fot\'on sobre la variedad  espacio temporal de Kerr, se emplea el uso de las t\'etradas, que es el
sistema utilizado por \cite{R10}, para describir las ecuaciones de movimiento  encontradas desde
el tensor de Riemann y de Ricci. Tambi\'en es usado, el Gauge  elegido por \cite{R10}, 
el cual aborda el problema relacionado con el horizonte de eventos del agujero negro. Un punto importante tiene que ver con la ecuaci\'on de Ernst 
debido a que desde all\'i se generan los factores de la m\'etrica y puesto  que las componentes del tensor de Ricci y de Riemann est\'an dadas por una combinaci\'on lineal del tensor de
Riemann, al anularse \'esta,  se obtienen las ecuaciones del espacio - tiempo axialmente sim\'etrico. Las ecuaciones de movimiento para el fot\'on en inmediaciones del agujero negro se escriben usando las  coordenadas de \textit{Boyer-Lindquist} (B.L), \cite{R6}, que permiten un cambio coordenado, para abordar este problema de una manera m\'as simple. Las coordenadas de B.L \cite{R6}, se usan particularmente porque minimizan el n\'umero de componentes fuera de la diagonal de la matriz y permiten establecer una diferencia entre el horizonte de eventos y la ergosfera. Adem\'as, se hace la separabilidad de Hamilton Jacobi para obtener las ecuaciones de movimiento del sistema, \cite{R12}.
Para las  soluciones de dichas ecuaciones de movimiento,  todas las integrales cu\'articas se han transformado a c\'ubicas
a partir del m\'etodo de Carlson, \cite{R1}. Se resuelven dichas  integrales (el\'ipticas) por los algoritmos del c\'odigo  implementado
y siguiendo este proceso para cada coordenada,  se encuentra la descripci\'on del comportamiento de los fotones en una regi\'on
conocida como ``puntos de retorno'' que es importante cuando se resuelve el problema del trazado de rayos.
Hay algunas caracter\'isticas  sobre el trazado de rayos, el cual simula el camino que siguen los fotones. Estos  pueden ir al infinito \'o
caer en el interior del agujero negro donde $R(r)=0$, entonces no hay ra\'ices reales,  y s\'i el punto de retorno es menor o igual
 al radio del horizonte de eventos, se considera  como si tuviera dos puntos de retorno,  el horizonte de eventos y el infinito (cuando el fot\'on cae en el interior del agujero).
 Finalmente, para la implementaci\'on del c\'odigo para el trazado de rayos, se encuentra que cada rutina del
 c\'odigo tiene una tarea espec\'ifica en resolver las ecuaciones de movimento en el sistema coordenado elegido por los m\'etodos 
 establecidos por \textit{Carter},  adem\'as,  expresa todas las coordenadas en funci\'on de un par\'ametro $p$ que corresponde a la terminaci\'on 
 de la geod\'esica cuando llega al observador y lo que sucede cuando hay una intersecci\'on de la geod\'esica con alguna superficie  de la regi\'on.
 As\'i mismo,  el estudio de  las geod\'esicas con los ejemplos ilustrados en este trabajo y teniendo en cuenta un factor importante
 que es tomar como variable independiente $r$ y $\mu$.
Se hace referencia tambi\'en al c\'alculo de las trayector\'ias de los fotones que se propagan desde el plano 
 ecuatorial en el infinito hasta el disco de acreci\'on que un observador lejano podr\'ia obtener con un telescopio. 
El color y la intensidad  se pueden relacionar con la energ\'ia recibida de  los fotones, y de esta manera generar una
imagen formada con  cada p\'ixel de las diferentes trayector\'ias, \cite{R6}. 
Las im\'agenes son tambi\'en distorcionadas por el efecto Doppler debido a la rotaci\'on, corrimiento al rojo (redshift) gravitacional y 
 doblamiento de la luz cerca al agujero negro, esto es una buena aproximaci\'on a lo que sucede en el espacio. Tambi\'en las 
 orientaciones que toman los fotones en la parte lejana del disco causan una
 asim\'etria en el disco generando un efecto de doblamiento en el observador.
 Este trabajo puede expandir sus horizontes a otro tipo de m\'etricas, y adem\'as se puede extender al problema de
 la transferencia radiativa. 

\subsection{La m\'etrica de Kerr}

La m\'etrica de Kerr  es una soluci\'on estacionaria y axialmente sim\'etrica a las ecuaciones de campo de Einstein. Para los espacios axialmente sim\'etricos se requiere que los coeficientes de la m\'etrica sean independientes de $x^0=t$ y $x^1$
adem\'as de las coordenadas de Boyer Lindquist, $(t,r,\theta,\phi)$. Una condici\'on importante es que el espacio-tiempo (E-T) debe ser invariante bajo  inversi\'on simult\'anea de $t$ y $\phi$, debido a que la fuente de campo gravitacional 
tiene movimiento rotacional alrededor del eje de sim\'etria. 
Esta inversi\'on es el difeomorfismo que preserva las relaciones de sim\'etria y el \'algebra tensorial, con estas condiciones la m\'etrica tiene  la forma
\begin{equation}
ds^2=e^{2\nu}(dt)^2-e^{2\Psi}(d\varphi-\omega dt)^2-e^{2\mu_{2}}(dx^{2})^2-e^{2\mu_{3}}(dx^{3})^2,
\label{g1}
\end{equation}
se deben tener en cuenta ciertas definiciones importantes para los t\'erminos relacionados con la m\'etrica,  una de ellas es que la m\'etrica 
es estacionaria y axialmente sim\'etrica, en general para el E-T  se tiene la ecuaci\'on del horizonte de eventos, la cual p\'ermite estudiar el 
problema en el plano ecuatorial
\begin{equation}
N(X^{2},X^{3})=0,
\label{f1}
\end{equation}
y esta debe  ser una superficie nula para que satisfaga las condiciones de este tipo de espacios. Usando el Gauge de libertad sugerido por \cite{R10}, se tiene  
\begin{equation}
e^{2(\mu_{3}-\mu_{2})}=\Delta(r),
\label{f4}
\end{equation}
con esta generalizaci\'on de la derivada covariante, teniendo en cuenta tambi\'en que para una superficie nula  $\Delta (r)=0$ y  junto con otras 
condiciones, se obtiene la soluci\'on para $ \Delta$
\begin{equation}
\Delta=r^{2}-2Mr+a^{2}.
\label{f9}
\end{equation}

La soluci\'on a la  ecuaci\'on (\ref{f9}) es de gran importancia en este trabajo porque de all\'i surge el an\'alisis para el movimiento del fot\'on y la descripci\'on para las ecuaciones de movimiento.

\subsection{Ecuaci\'on de Ernst y la m\'etrica de Kerr}
La ecuaci\'on de Ernst permite una soluci\'on de la forma
\begin{equation}
 \widetilde{E}=-p\eta-iq\mu,
 \label{f11}
\end{equation}
que sirve para encontrar los coeficientes de la m\'etrica
\begin{equation}
p^{2}+q^{2}=1,
\label{f12}
\end{equation}
siendo $p$ y $q$ constantes, que ser\'an representadas en este trabajo como $a$ y $M$, luego se entrar\'a  en detalle para cada una de estas.
despu\'es del c\'alculo algebraico, a partir de la soluci\'on a la ecuaci\'on de Ernst, se obtiene lo siguiente
\begin{equation}
 \Sigma^{2} = (r^{2}+a^{2})^{2} - a^{2}\Delta \delta,
 \label{f37}
\end{equation}
\begin{equation}
 e^{2\Psi}= \frac{\delta\Sigma^{2}}{\rho^{2}},
 \label{f38}
\end{equation}
\begin{equation}
 \omega= \frac{2aMr}{\Sigma^{2}},
 \label{f39}
\end{equation}
\begin{equation}
 e^{2\nu} = \frac{\rho^{2}\Delta}{\Sigma^{2}}.
 \label{f41}
\end{equation}
Las componentes del tensor de Ricci y de Einstein, se encuentran dadas por una combinaci\'on lineal de las componentes del tensor de Riemann, que 
al anularse,  se obtiene la ecuaci\'on para el espacio tiempo axialmente sim\'etrico y estacionario. De este c\'alculo matem\'atico se obtiene
\begin{equation}
e^{\mu_{3}+\mu_{2}=\frac{\rho^{2}}{\sqrt{\Delta}}},
\label{f62}
\end{equation}
donde el  gauge elegido fue usado por  \cite{R10}, el cual viene dado como  $e^{\mu_{3}-\mu_{2}=\frac{\rho^{2}}{\sqrt{\Delta}}}$. Al generar las soluciones para $\mu_{3}$ y $\mu_{2}$ se tiene
\begin{eqnarray}
e^{2\mu_{2}}=\frac{\rho^{2}}{\Delta}, \label{f63}
e^{2\mu_{3}}=\rho^{2}, \label{f64}
\end{eqnarray}
con esto se completa la soluci\'on para los coeficientes de la m\'etrica, que al sustituir en la m\'etrica, esta  se reescribe como
\begin{equation}
ds^{2}=\rho^{2}\frac{\Delta}{\Sigma^{2}}(dt^{2})-\frac{\Sigma^{2}}{\rho^{2}}(d\varphi- \frac{2aMr}{\Sigma^{2}}dt)^{2}sen^{2}\theta - \frac{\rho^{2}}{\Delta}(dr)^{2}-\rho^{2}(d\theta)^{2}.
\label{f65}
\end{equation}

\subsection{Las ecuaciones de movimiento del fot\'on}

A partir de la ecuaci\'on (\ref{g1})  y reemplazando las condiciones dadas por el Gauge de libertad \cite{R10}, la m\'etrica queda finalmente expresada de la siguiente forma

\begin{eqnarray}
 ds^{2}=-\left(1-\frac{2Mr}{\Sigma} \right) dt^{2} - \left(\frac{4aMrsen^{2} \theta}{\Sigma} \right) dt d\varphi + \label{i2} \nonumber \\
 \left(\frac{\Sigma}{\Delta}\right)dr^{2} + \Sigma d\theta^{2} + \left(\frac{r^{2}+ a^{2}+2Ma^{2}rsen^{2}\theta}{\Sigma} \right) sen^{2}\theta d\varphi^{2},
 \end{eqnarray}
 
siendo  $M$  la masa del agujero negro, $a$ el par\'ametro de rotaci\'on relacionado con el momentum angular $(0 \leq a \leq M)$, 
que diferencia la m\'etrica de Kerr con la de Schwarszchild y debido a las siguientes condiciones que son de gran importancia para los coeficientes de la m\'etrica,

\begin{eqnarray}
 \Sigma&=&\rho^{2},\label{condici1}\\
 \Delta&=& r^{2}-2Mr+a^{2},\label{condici2}\\
 \Sigma&=&r^{2}+a^{2}\cos^{2}\theta,\label{condici3}\\
 \Sigma^{2}&=&(r^{2}+a^{2})^{2}-a^{2}\Delta \sin^{2}\theta,\label{condici4}\\
 \mu^{2}&=&\cos^{2}\theta,\label{condici5}\\
 \mu^{2}&=&1-\sin^{2}\theta,\label{condici6}\\
 \sin^{2}\theta&=&1-\mu^{2},\label{condici7}\\
 \sin^{2}\theta&=&\delta,\label{condici8}\\
 \rho^{2}&=&r^{2}+a^{2}\mu^{2}\label{condici9}.
 \end{eqnarray}
 
 Que son las expresiones \'utiles para encontrar las ecuaciones de movimiento del fot\'on, donde $\sigma$ es el tiempo propio de la part\'icula, $\mu$ representa las distintas \'orbitas de los
 fotones, $a$ el par\'ametro de rotaci\'on $r,\theta$ las coordenadas espaciales y $\rho$ el par\'ametro de ajuste. La representaci\'on de  la m\'etrica (ver ecuaci\'on (\ref{i2})) en t\'erminos del horizonte de eventos que matem\'aticamente est\'a  localizado por fuera de la ra\'iz en la ecuaci\'on (\ref{f9}) para $\Delta=0$
\begin{equation}
r=r_{+}\equiv M \pm \sqrt{(M^{2}- a^{2})},
 \label{i3}
\end{equation}
al usar la condici\'on expresada en la ecuaci\'on (\ref{condici2}) y el signo $\pm$ determina la regi\'on a estudiar, se obtiene
\begin{equation}
0=r^{2}-2Mr+a^{2}.
 \label{i4}
\end{equation}
Debido a que $(0\leq a \leq M)$, $r$ solo  puede tomar los valores que surgen a partir de la ecuaci\'on (\ref{i4}) y que en $a=0$ representa soluci\'on  de {\textit{Schwarszchild}}  
\[
\left\{ \begin{array}{lcl}
\mbox{si},a=0 & r=2M\\
\mbox{si}, a=M & r=M,
\end{array}
\right.\nonumber
\]
el l\'imite est\'atico que se encuentra  fuera de la  frontera del agujero negro en la ergosfera, se determina por la ra\'iz de $\Sigma-2Mr=0$, y se obtiene a partir
de las condiciones establecidas en las ecuaciones (\ref{condici2}) y (\ref{condici4}). Como $\Delta=0$, entonces al tomar la ecuaci\'on (\ref{i4}) completar el cuadrado
e igualar con (\ref{condici2}) y (\ref{condici3}) se tiene
\begin{equation}
 \Sigma -2Mr=0
 \label{i7}
\end{equation}

el c\'alculo de $\Sigma$ es importante porque es un par\'ametro que aparece en la m\'etrica. Desde la ecuaci\'on (\ref{i3}) se obtiene el horizonte de la ergosfera. Para un observador que sigue la l\'inea como de tiempo,  debe ser arrastrado en la direcci\'on positiva de $\Phi$. Si esta en el  interior del l\'imite est\'atico (ergosfera), estos observadores tienen acceso a una energ\'ia de trayector\'ia 
negativa la cual en principio,  extrae energ\'ia del agujero negro,  \cite{R3}, \cite{R2}.

\subsection{Separabilidad de Hamilton Jacobi}

La separabilidad de Hamilton Jacobi se aplica para tener las ecuaciones de movimiento del fot\'on y con  la ecuaci\'on de las geod\'esicas 

\begin{equation}
 \frac{d^{2}x^{\alpha}}{d \sigma^{2}}+{\Gamma}^{\alpha}_{\mu \nu} u^{\mu} u^{\nu}=0
 \label{i10},
\end{equation}

donde $\sigma$ representa el tiempo propio de la part\'icula y el par\'ametro af\'in para los fotones, ver \cite{R1}, $u^{\nu}$ es la cuadri-velocidad 
y ${\Gamma}^{\alpha}_{\mu \nu}$ son los coeficientes de conexi\'on que toman la forma de la ecuaci\'on de la geod\'esica para aplicarle la separabilidad

\begin{equation}
 2\frac{dS}{d\sigma}=g^{ij}\frac{\partial S}{\partial x^{i}}\frac{\partial S}{\partial x^{j}},
 \label{i11}
\end{equation} 
con  $S$ como la acci\'on cl\'asica que despu\'es de algunos c\'alculos matem\'aticos, se obtienen
un par de ecuaciones que se expresan en t\'erminos de $R$ y $\Theta$, se reconocen como los par\'ametros, en los cuales se desea hacer la separaci\'on y $Q$ que representa la constante de separaci\'on de
Carter. De esta forma, se encuentra la soluci\'on para la acci\'on 
\begin{equation}
\int^{r}{\frac{dr}{\sqrt{R}}}= \int^{\theta}\frac{d\theta}{\sqrt{\Theta}}.
\label{i24}
\end{equation}

De la ecuaci\'on (\ref{i40}) depende el planteamiento de las ecuaciones de movimiento para el fot\'on y del m\'etodo de Carter 
que es donde se empiezan a implementar los m\'etodos num\'ericos para resolver estas ecuaciones, \cite{R1}. 

\subsection{Cuadri-momento del fot\'on}

Las \'orbitas de movimiento son descritas por las constantes de movimiento de Carter

\begin{equation}
 Q= k-(L_{z}-aE)^{2},
 \label{i31}
\end{equation}

en la cual $Q$ es la constante de Carter $L_{z}$ el momento angular sobre el eje de rotaci\'on y $E$ la energ\'ia. De la ecuaci\'on (\ref{i31}) se obtiene el cuadri-momento del fot\'on  

\begin{eqnarray}
 \Sigma\frac{dr}{d\sigma}&=& \pm \sqrt{R}, \label{i40} \nonumber \\
 \Sigma\frac{d\theta}{d\sigma}&=& \pm \sqrt{\Theta}, \nonumber \\
 \Sigma\frac{d\varphi}{d\sigma}&=& -\left(a-\frac{\lambda}{\sin^{2}\theta} \right) +a\frac{T}{\Delta}, \nonumber \\
 \Sigma\frac{dt}{d\sigma}&=& -a(a \sin^{2} \theta-\lambda)+(r^{2}+a^{2})\frac{T}{\Delta}, 
\end{eqnarray}

que es el conjunto de ecuaciones que describe la trayector\'ia del fot\'on en t\'erminos de cada coordenada y que en otras palabras se pueden reconocer
como el resultado de la separaci\'on. Estas ecuaciones son solucionadas en la siguiente secci\'on y para ello se necesita ejecutar el programa,  \cite{R2}.

\subsection{Soluci\'on a las ecuaciones de Movimiento}

Ahora se  expresan  las coordenas de B.L  en t\'erminos de una transformaci\'on $p$. Las funciones $R$ y $\Theta$
no deben ser negativas y los puntos de frontera de esta regi\'on ser\'an puntos de retorno adem\'as  las coordenadas para un fot\'on emitido son  $r_{ini}$ y $\theta_{ini}$. El movimiento 
esta confinado entre dos puntos de retorno $r_{tp1}$ y $r_{tp2}$, siendo una coordenada radial. En el caso $\theta_{tp1}$ y $\theta_{tp2}$ se representa por una 
coordenada poloidal. El movimiento de un  fot\'on puede no necesariamente estar determinado por una frontera, esto significa que este puede ir al infinito \'o caer al interior del agujero negro,  el \'ultimo caso, particularmente hace referencia a $R(r) =0$, donde no hay ra\'ices reales. 
En un movimiento poloidal, es importante tener en cuenta que $\theta=0$ y $\theta=\pi$. Por ejemplo, un fot\'on con $\lambda=0$ se dirige 
por el eje de esp\'in debido al momentum angular nulo, y los cambios de signo en la velocidad angular $d\theta/d\sigma$.
La coordenada azimutal varia de  $\varphi$ a $\varphi \pm \pi$,  lo que implica que el eje de esp\'in no es un punto de retorno ya que no son ra\'ices reales de las ecuaciones $\Theta=0$, \cite{R2,R1}.
Por \'ultimo, usando las  Integrales el\'ipticas Weierstrass-Jacobi, que solucionan num\'ericamente las ecuaciones de movimiento en el problema de la trayector\'ia de los fotones sobre la geod\'esica, 
debido a que surgen polinomios de grado superior. Para m\'as informaci\'on sobre las integrales el\'ipticas ver \cite{R7}.

\subsection{El m\'etodo de Carlson}

Este m\'etodo es usado para escribir las coordenadas  en t\'erminos de una nueva funci\'on $p$, tal como se enunci\'o en la secci\'on anterior, esta 
nueva coordenada implica  resolver cada coordenada para  las ecuaciones de movimiento. Principalmente es una representaci\'on de 
para la soluci\'on de integrales el\'ipticas ver \cite{R1}, 

\begin{equation}
 p=\pm \int^{r}{\frac{dr}{\sqrt{R}}}= \pm \int^{\theta}{\frac{d\theta}{\sqrt{\Theta}}},
 \label{b4}
\end{equation}

donde $p$, es el par\'ametro que describe el movimiento del fot\'on a lo largo de la geod\'esica para
cada coordenada. Gracias al par\'ametro $p$   se obtienen ecuaciones reducidas a polinomios c\'ubicos y son las que resuelven el programa a partir de las integrales el\'ipticas. \cite{R2}\\

Esto se debe, a que se quiere hallar la posici\'on original de donde se emiti\'o el fot\'on \'o los caminos que tomo este, por lo tanto  se hace el trazado desde el observador al emisor, a lo largo de geod\'esicas y con  al m\'etodo de Newton
Raphson se traza cada coordenada que  depende de la cantidad y tipo de ra\'iz que tenga al solucionar las ecuaciones de movimiento.

\begin{itemize}
\item \textit{Coordenada $\mu$}.
\item \textit{Coordenada $r$}. 
\item \textit{Coordenada t},  $\Phi$  y par\'ametro af\'in $\sigma$\\
\end{itemize}

Las anteriores son las coordenadas de movimiento que se simulan con el c\'odigo.

\section{El C\'odigo}

El c\'odigo \textit{YNOGK-GEOKERR}   est\'a escrito en \textit{Fortran95} y fue  desarrollado por \cite{R1}, posteriormente modificado por \cite{R2}. El c\'odigo resuelve las ecuaciones de movimiento para las geod\'esicas nulas  en un espacio axialmente sim\'etrico y en base a estos resultados simula un trazado de rayos con diferentes condiciones, computando las trayector\'ias que sigue el fot\'on sobre la variedad espacio temporal.

Los m\'etodos usados en el c\'odigo son los empleados por Carter para resolver ecuaciones el\'ipticas  de primer y segundo tipo. 
Los datos obtenidos en este c\'odigo se reproducen en una grilla con un par\'ametro de imp\'acto determinado por el programa y que el usuario
puede ajustar, as\'i mismo, las otras componentes del c\'odigo se usan  para estudiar el comportamiento de las geod\'esicas y resolver ecuaciones cu\'articas con par\'ametro $p$.  Tambi\'en,  las integrales el\'ipticas se transforman a c\'ubicas. De esta manera las cuatro coordenadas y el par\'ametro af\'in se expresan en funci\'on de $p$. 
El inter\'es de realizar esto, es determinar la posici\'on original donde 
el fot\'on fue emitido o el camino seguido por este. Para que el c\'alculo sea m\'as preciso los fotones se trazar\'an  hacia atras, desde el observador al emisor siguiendo geod\'esicas.
Pero no todos los fotones que empiezan desde un observador dado van a la regi\'on de emisi\'on, puede que algunos se pierdan en el camino,  cuando los fotones alcanzan el infinito o caen en el horizonte de eventos.

Ad\'emas el c\'odigo usa  la rutina  de transferencia radiativa de \textit{Monte-Carlo}, que requiere hacer unas transformaciones entre los marcos de 
 referencia del emisor y el sistema de B.L  donde se puede observar que la emisi\'on en el marco de referencia del emisor es isotr\'opica, como se observa en la figura \ref{fig_Rayos}, pero desde el sistema B-L es anisotr\'opico, debido al efecto de apantallamiento Doppler.
 Este c\'odigo \cite{R2},  es una modificaci\'on de \cite{R1},  pero logra una  velocidad de procesamiento de los datos mayor, adem\'as se incluyen unas rutinas  nuevas para la transferencia radiativa.

\section{Resultados}

\subsection{Rayos}
En la figura \ref{fig_Rayos} se muestra los rayos de un grupo de geod\'esicas  nulas emitidas isotr\'opicamente por una part\'icula moviendose alrededor del agujero negro en una \'orbita circular estable $r_{ms}$. 

\begin{figure}[h!]
\centering
\includegraphics[width=0.5\textwidth]{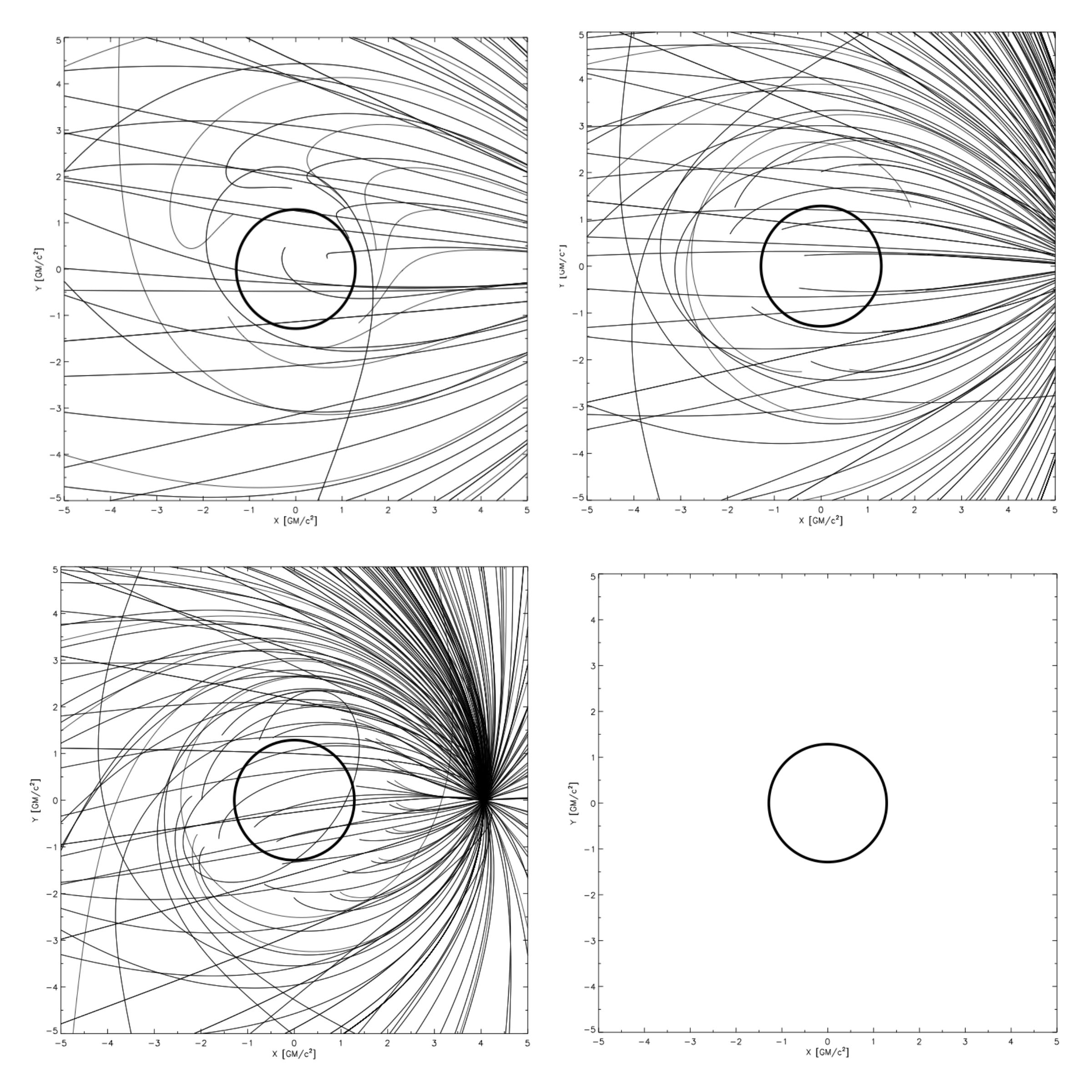}
\caption{En esta simulaci\'on se tomaron diferentes valores para $a$ (momentum angular de esp\'in) $a=-0.9375, 0.5,0$ y $1$ respectivamente. La escala de grises  representa el par\'ametro af\'in}
\label{fig_Rayos}
\end{figure}
\begin{figure}[h!]
\centering
\includegraphics[width=0.5\textwidth]{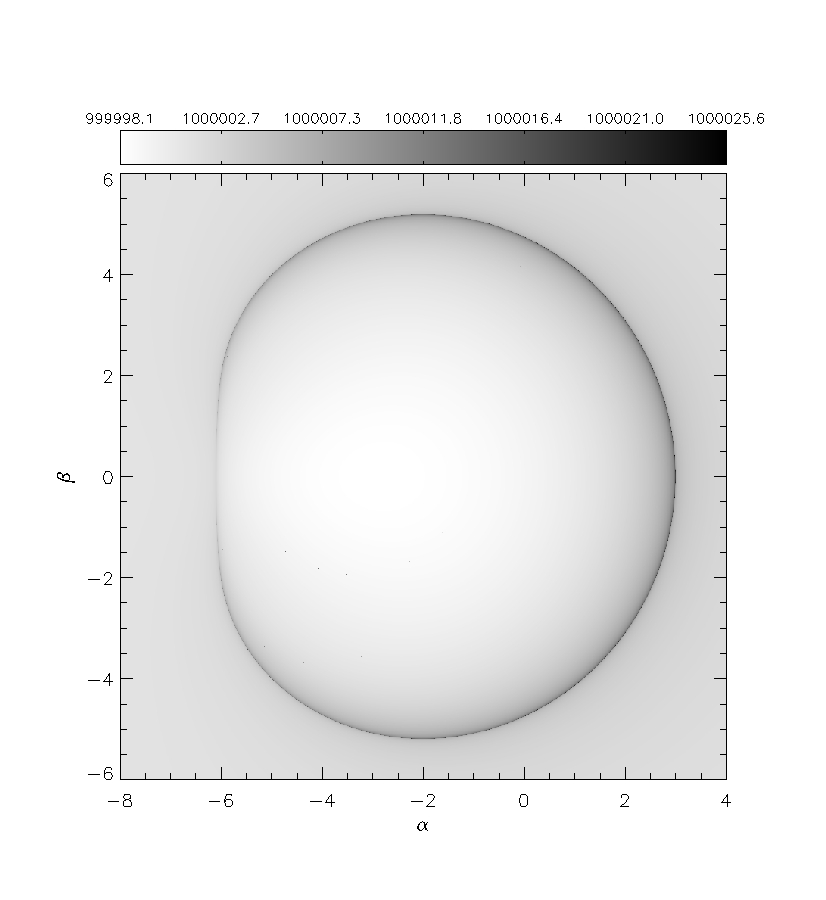}
\caption{La intensidad del color representa el par\'ametro af\'in, con momentum angular de esp\'in $a=0.998$ y $\alpha$, $\beta$ representan el par\'ametro de impacto.}
\label{fig_shadow}
\end{figure}
\begin{figure}[h!]
\centering
\includegraphics[width=0.5\textwidth]{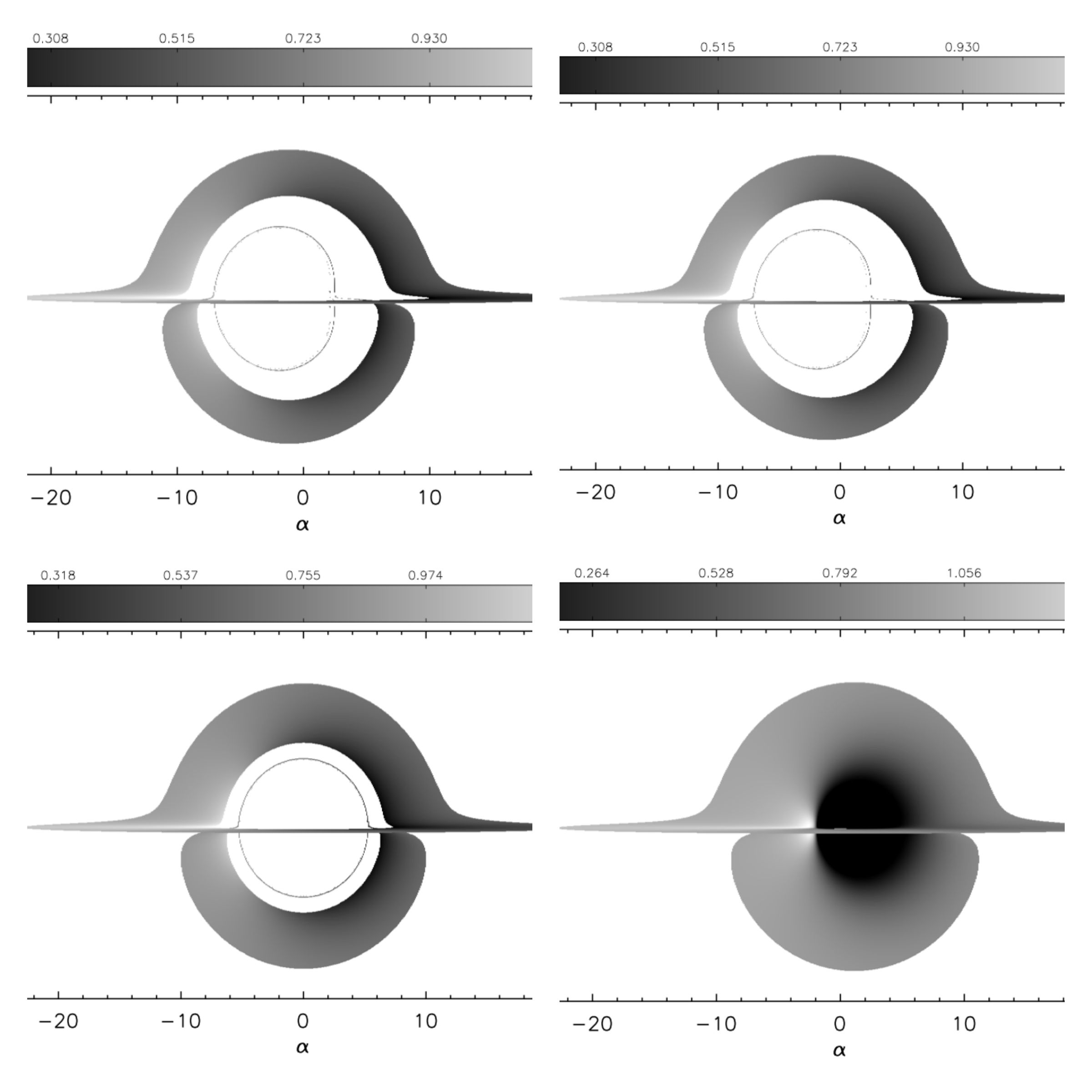}
\caption{En esta simulaci\'on se tomaron distintos valores para $a=-1,-0.998,0$ y $1$ donde se ve lo \'unico que cambia es $r$. La escala de grises representa el redshift gravitacional al igual que el ejemplo anterior los ejes $\alpha$ y $\beta$ son los par\'ametros de impacto que describen el tama\~no y la posici\'on de la imagen.}
\label{fig_Diskm }
\end{figure}
En la figura \ref{fig_Rayos}, el agujero negro esta situado en origen coordenado y alrededor se observan las diferentes trayector\'ias que siguen los fotones, con distintos valores para el momento angular. y los $x$ e $y$ son una especie de coordenadas cartesianas en el plano ecuatorial del agujero negro, esta 
figura  es muy importante porque  se puede apreciar el efecto de lente gravitacional,  \cite{R14}.
\begin{figure}[h!]
\centering
\includegraphics[width=0.5\textwidth]{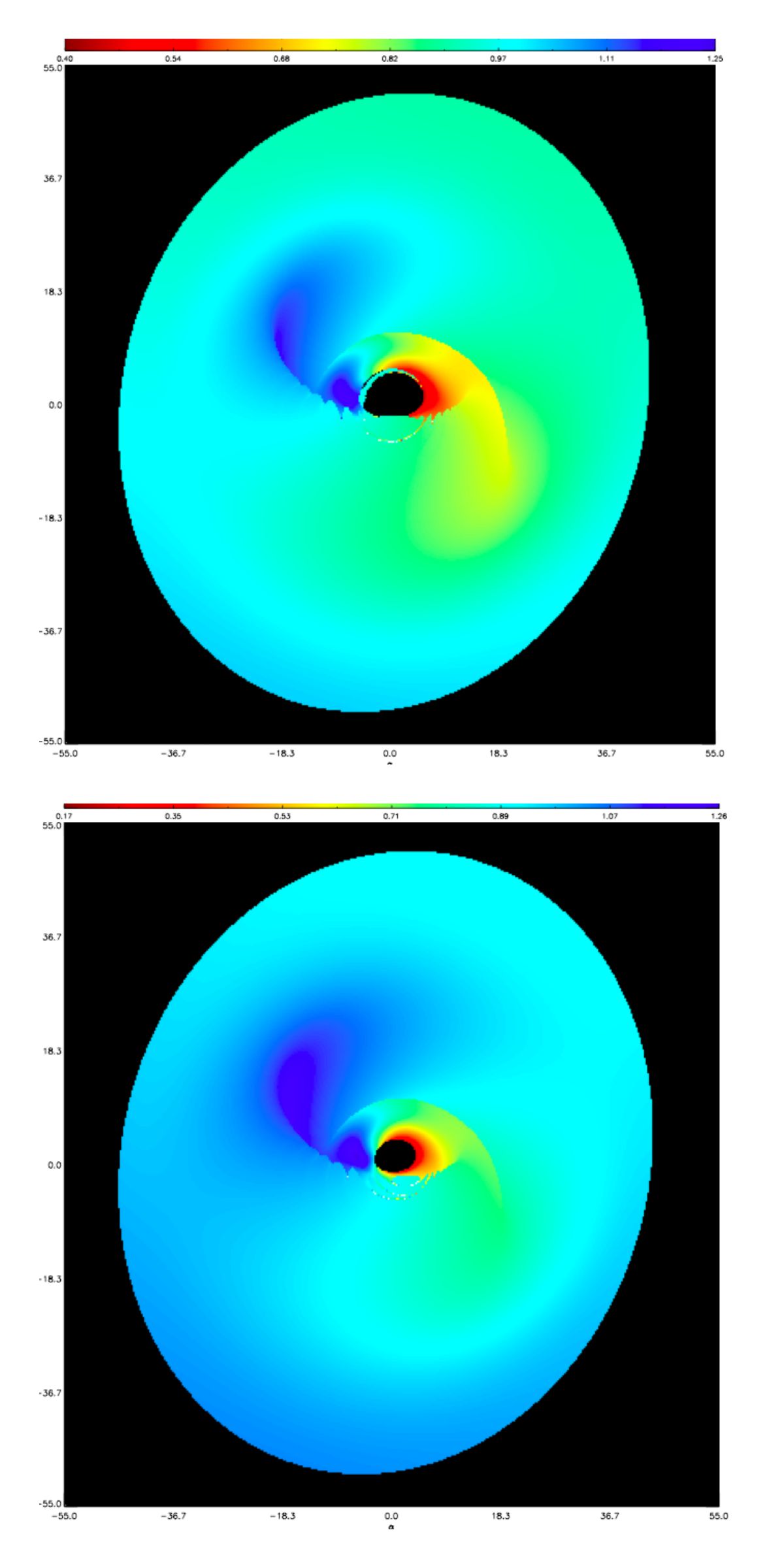}
\caption{La intensidad de color se torna mas intensa debido al efecto Doppler gravita\-cional, si se modifica el \'angulo desde el observador el disco se deforma. Los ejes representan el p\'arametro
de impacto sobre la grilla.}
\label{fig_warpm}
\end{figure}

\subsection{Sombra del agujero (Shadow)}

La figura \ref{fig_shadow}, muestra la imagen de la sombra del agujero negro, en el cual la intensidad de  grises es representada por el valor del par\'ametro  af\'in $\sigma$ que mide el observador en la posici\'on final, y en funci\'on de  $\sigma(p)$ el momento angular del agujero, en este caso es de $a=0.998$ y la escala de grises representa el valor de  $\sigma$  evaluado para un observador en una posici\'on final cualquiera con respecto al agujero negro, o apareciendo en el radio inicial. Los ejes $\alpha$ y $\beta$ son los par\'ametros de impacto que describen el tama\~no y la posici\'on de la imagen.

\subsection{Discos de acreci\'on y discos deformados (Warp)}

En la figura  \ref{fig_Diskm } se muestra  los discos de acreci\'on.  Se  calcula los perfiles de emisi\'on para $FeK_{\alpha}$,  \cite{R1} y su espectro,  \cite{R5}. Usualmente los colores de la imagen representan el redshift o el flujo de intensidad para el observador, de la emisi\'on proveniente 
del disco. La inclinaci\'on del disco es $\theta_{obs}=85^{o}$ y la distancia del observador es $40r_{g}$. Se puede ver la parte interior 
del agujero negro y la parte inferior del disco.

La acreci\'on en los agujeros negros, ver figura \ref{fig_warpm}, es importante  porque  se puede considerar como una forma de transformar energ\'ia gravitacional en radiaci\'on
y que generalmente ocurre en un disco, su forma cambia dependiendo de diferentes cantidades, por ejemplo el \'angulo en el cual se encuentra
el observador, que se relaciona con el transporte de momentum angular de las regiones interiores a las regiones exteriores del sistema, eso define
de forma general su morfolog\'ia.

\section{Conclusiones}

La importancia en el desarrollo de herramientas computacionales  ha permitido resolver y simular, diversas situaciones adem\'as 
de abordar distintos problemas que con ayuda del c\'alculo num\'erico, que permiten un trabajo m\'as efectivo y que lleva los problemas a un punto donde
se puede aproximar en gran instancia a la realidad. Este trabajo se basa principalmente
en dos c\'odigos que describen los espacios axialmente sim\'etricos YNOGK, \cite{R2} y GEOKERR, \cite{R1}. El segundo de gran importancia porque es donde surge esta laboriosa tarea a partir de la m\'etrica establecida por Kerr en 1960 y que por mas de 50 a\~nos astrof\'isicos  han estudiado.  
La separaci\'n de Hamilton-Jacobi  usa las cuatro coordenas y el par\'ametro af\'in, para atacar el problema del trazado de rayos. Gracias
a esto se logra determinar las orbitas de los fotones alrededor del agujero negro y la trayector\'ia medida desde el observador. Encontrando las soluciones
a las ecuaciones de movimiento cuyas ra\'ices especifican
el comportamiento de los fotones.  Cada rutina se encarga de resolver num\'ericamente 
sus respectivas coordenadas, una tarea compleja de realizar manualmente debido a que surgen ecuaciones no lineales y  en su mayor\'ia se resuleven usando el
m\'etodo de Newthon-Raphson.

Finalmente con este trabajo, se pretende llevar a otro tipo de m\'etricas y estudiar diferentes problemas relacionados con los agujeros negros, adem\'as
de estudiar problemas como lensing gravitacional y transferencia radiativa, entre otros.

\renewcommand{\refname}{Bibliograf\'ia}
\bibliographystyle{harvard}
\bibliography{Jonathan}

\end{document}